\documentclass[aps,prb, twocolumn]{revtex4-1}

\usepackage{amsmath}
\usepackage{comment}
\usepackage[utf8]{inputenc}

\usepackage{graphicx} 
\usepackage{epstopdf} 
\usepackage{color}

\begin{document}

\preprint{Version 2}
 
\title{Phase diagram of $^{4}$He on graphene}

\author{Jodok Happacher}
\affiliation{Department of Physics, Arnold Sommerfeld Center for Theoretical Physics and Center for NanoScience, University of Munich, Theresienstrasse 37, 80333 Munich, Germany}
\author{Philippe Corboz}
\affiliation{Theoretische Physik, ETH Zurich, CH-8093  Zurich, Switzerland}
\author{Massimo Boninsegni}
\affiliation{Department of Physics, University of Alberta, Edmonton, Alberta, Canada, T6G 2G7}
\author{Lode Pollet}
\affiliation{Department of Physics, Arnold Sommerfeld Center for Theoretical Physics and Center for NanoScience, University of Munich, Theresienstrasse 37, 80333 Munich, Germany }
\email[]{Lode.Pollet@physik.uni-muenchen.de}

\date{\today}

\begin{abstract}

The low temperature phase diagram of $^4$He adsorbed on a single graphene sheet is studied by  
computer simulations of a system consisting of nearly a thousand helium atoms. 
In the first layer two commensurate solid phases are observed with fillings $1/3$ and $7/16$, respectively, separated by a domain wall phase, as well as an incommensurate crystal at a higher coverage. No evidence of a thermodynamically stable superfliuid phase is found for the first adlayer.
Second layer promotion occurs at a coverage  of 0.111(4)~\AA$^{-2}$. 
In the second layer  two  phases are observed, namely a superfluid and an incommensurate solid, with no  commensurate solid  intervening between these two phases.  The computed phase diagram closely resembles that  predicted for helium on graphite.

\end{abstract}


\maketitle

\section{\label{Introduction}Introduction}
The study of films of highly quantal fluids, such as helium, is motivated by the search for novel phases of matter in confinement and/or reduced dimensions.
Indeed, an experimentally controllable way of making quasi-2d interacting $^4$He systems is to adsorb  a thin film of $^4$He on a substrate. On weakly attractive substrates, such as those of some alkali metals, $^4$He forms superfluid  films (down to a monolayer \cite{bon99} thin) whose thickness smoothly increases with chemical potential, with no evidence of layering.\cite{bon03,vancleve08} On the other hand, on stronger substrates adsorption occurs through the formation of successive, well-defined layers (up to seven on graphite), with essentially no quantum-mechanical atomic exchanges taking place between the first few adjacent adlayers. The phase diagram on these substrates is richer, displaying a variety of phases, including crystalline ones, either commensurate or incommensurate with the underlying substrate. 
\\ \indent
Crowell and Reppy~\cite{Crowell} raised some time ago the possibility of a ``supersolid" phase,\cite{Django} characterized by simultaneous density and superfluid long-range order, in the vicinity of a possible crystalline phase of the second helium adlayer, registered with the underlying graphite substrate. This contention  has recently been reiterated.~\cite{fuku} The most reliable, first-principle numerical studies of
helium films on graphite have yielded no evidence of such a phase, as no registered crystal is observed in the second adlayer.~\cite{Corboz} 
\\
\indent
Graphene (a single sheet of graphite) has also been theoretically considered  as a possible substrate for helium adsorption, and the phase diagram of the adsorbate in the low temperature (i.e., $T\to 0$) limit has been computed by means of different numerical techniques.\cite{Gordillo1st,Cazorla,Gordillo2nd,Kwon} A single sheet of carbon atoms is somewhat less attractive than a graphite substrate. Quantitatively, the atomic binding energy for the first $^4$He adlayer is reduced by approximately 10\% (about 13.4 K) compared to graphite. \cite{Gordillo1st} One might imagine that the energy offset could lead to different physical behavior, but first principle calculations suggest
that a difference of that order of magnitude in the adsorption potential is likely to have little or no  effect on the phase diagram of helium on graphite.~\cite{Corboz} Thus, one may expect no qualititative differences in the phase diagram of $^4$He adsorbed on graphene versus that on graphite. 
Indeed, that is the conclusion at which most numerical studies carried out so far for this system have arrived, with an outstanding puzzle concerning a possible superfluid response in the first adsorbed layer, near and at commensurate filling, reported in Refs. \onlinecite{Cazorla,Kwon}.
\\ \indent
We report in this article results of a theoretical study of the low temperature phase diagram of $^4$He adsorbed on graphene, based on computer simulations. We consider here a system comprising at least twice as many $^4$He atoms than in previous studies by others, the goal being that of attempting a reliable extrapolation of the physics of the system in the thermodynamic limit. The main findings of our study are largely in line with most previous works on graphene, but with no evidence of any ``supersolid'' phase, neither in the second nor in the first adsorbed layer.
\\
\indent
The remainder of this paper is organized as follows: first, we briefly discuss the model of the system that is used, as well as the computational methodology; then, we proceed with the illustration of the results, focusing on the first and second adlayers. We summarize our main results in the Conclusions, where we also address issues that may be the subject of future work.

\section{\label{Methodology}Methodology}
In order to study numerically  the physical properties of $^4$He on a graphene sheet,  we performed equilibrium, large-scale computer simulations of a model of the system of interest, using the continuous-space Worm Algorithm.~\cite{WormAlgoshort,WormAlgolong}  Graphene is modeled as an ideal, two-dimensional honeycomb lattice, with a carbon-carbon bond length $a$=1.42 \AA. Carbon atoms are treated as fixed particles in our simulation, an assumption justified by their relatively large mass, compared to that of the $^4$He atoms. The system is enclosed in a three-dimensional cell, shaped as a parallelepiped; the graphene sheet is aligned parallel to the $xy$ plane (at $z$=0). Periodic boundary conditions are used in all directions, but the box is sufficiently elongated in the $z$ direction to make the boundary condition immaterial. 
The ensuing, quantum-mechanical many-body Hamiltonian is the following:
\begin{equation}
\hat H=-\frac{\hbar^2}{2m}\sum_i \nabla_i^2+\sum_{i<j}v(r_{ij})+\sum_{iL}u(|{\bf r}_i-{\bf R}_L|)
\label{ham}
\end{equation}
where $m$ is the mass of the $^4$He atoms, $v$ is a pairwise potential describing the interaction between two helium atoms, whereas $v$ describes the interaction of each helium-carbon pair, with ${\bf{R}}_L$ the position of a carbon atom. Both $u$ and $v$ are assumed to depend only on the relative distance between two particles.
The interaction between  helium atoms $v$ is taken to be the accepted Aziz pair potential, \cite{Aziz} whereas for the carbon-helium potential $u$ a Lennard-Jones potential is used, with parameters $\epsilon=16.2463$K and $\sigma=2.74 \AA$  chosen following Ref.~\onlinecite{CarlosPotential}. 
The simulated systems comprise close to one thousand  $^4$He atoms. The largest honeycomb lattice simulated here has size 49.19~\AA $\times$ 51.12~\AA, and consists of 960 carbon atoms. 
All the simulations are performed in the grand canonical ensemble,  at finite temperatures ranging from 0.5 K to 1.0 K. The results at the two different temperature differ very little, suggesting that they are essentially ground state estimates, at least as far as energetics and structures are concerned. Finally, all the results presented here are independent of the initial atomic configuration utilized in the simulation.

\section{\label{FirstLayer}Results for the first Layer}

\begin{figure}
\includegraphics[angle={-90},width=1.0\linewidth]{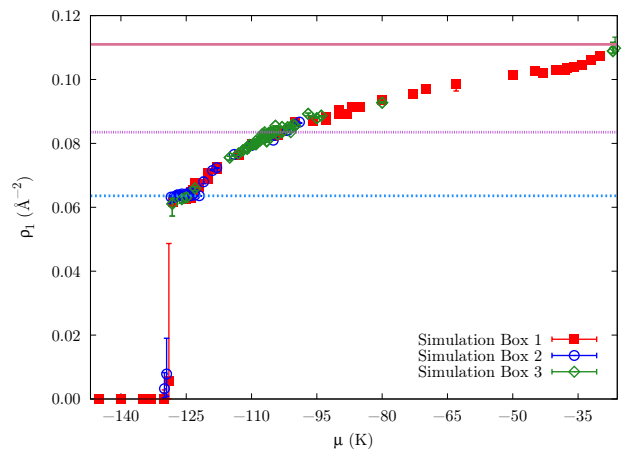}
\caption{\label{fig:dens_mu_layer1}(Color online). Average coverage versus chemical potential for the first $^4$He adlayer on graphene at a temperature $T=0.5$K. Data are shown for different system sizes, namely: Simulation Box 1, $49.19\AA \times 42.60\AA$ (commensurate with an ideal graphene lattice), Simulation Box 2, $51.65 \AA \times 46.86 \AA$ (commensurate with the C1/3 phase), and Simulation Box 3, $49.19 \AA \times 51.12 \AA$ (commensurate with the C7/16 phase). Results are independent of the initial configuration of the atoms. The first layer becomes populated around $\mu=- 130$\ K, forming a C1/3 crystal phase. The horizontal lines denote the density corresponding to maximum first layer occupation, the C7/16 phase and the C1/3 phase, top to bottom.
}
\end{figure}

We start the discussion of our results with the first layer. Fig.\ref{fig:dens_mu_layer1} shows the computed average  equilibrium $^4$He coverage (two-dimensional density) as a function of chemical potential $\mu$. 

\begin{table}[h]
 \caption{Winding number squared along $x$ and $y$ axis for $8 \times 6$ $^4$He atoms in the C1/3 phase.}
 \begin{ruledtabular}
 \begin{tabular}{|c|c|c|}
   $T [K]$    & $\langle W_x^2 \rangle$  & $\langle W_y^2 \rangle$ \\
  \hline 1.0  &$6(5) \times 10^{-5}$ & $6(4) \times 10^{-5}$ \\
  0.1		   & 	$4(3) \times 10^{-4}$ & $8(3) \times 10^{-3}$ \\
  0.05	   &   $3(3) \times 10^{-4}$ & $4(3) \times 10^{-3}$ \\
  0.02	   &	$9(6) \times 10^{-3}$ & $6(4) \times 10^{-3} $\\
 0.01	   &	$0.028(12)$		& $0.036(13)$ \\ 
 \end{tabular}
 \end{ruledtabular}
  \label{table:8x6}
 \end{table}

\begin{table}[h]
 \caption{Winding number squared along $x$ and $y$ axis for $8 \times 8$ $^4$He atoms in the C1/3 phase.}
 \begin{ruledtabular}
 \begin{tabular}{|c|c|c|}
   $T [K]$    & $\langle W_x^2 \rangle$  & $\langle W_y^2 \rangle$ \\
  \hline 1.0  &$2(2) \times 10^{-5}$ & $0.0(0)$  \\
  0.1		   & 	$0(0) $ & $2(2) \times 10^{-4}$ \\
  0.05	   &   $3(3) \times 10^{-4}$ & $0.0(0)$ \\
  0.02	   &	$0.0(0) $ & $0.0(0)$\\
 0.01	   &	$0.0(0)$		& $0.0(0)$ \\ 
 \end{tabular}
 \end{ruledtabular}
  \label{table:8x8}
 \end{table}

\begin{figure}
\includegraphics[angle={-90},width=1.0\linewidth]{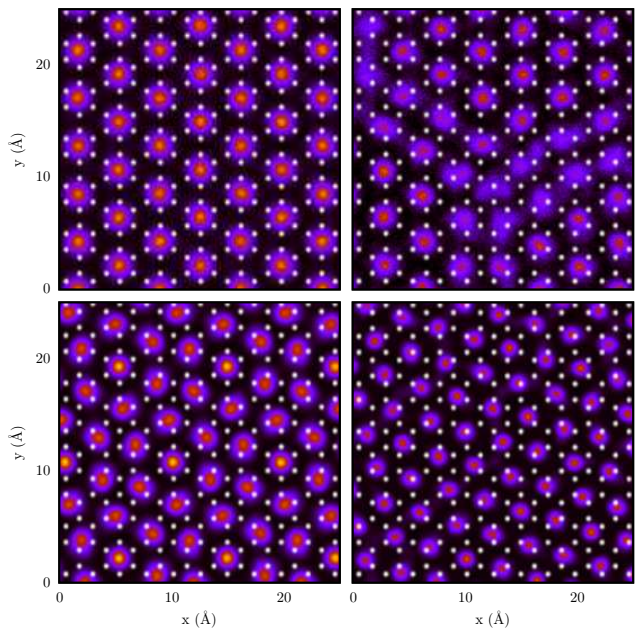}
\caption{\label{PicC7/16}(Color online). {\it Top Row: Left:} The commensurate C1/3 phase. {\it Top Row: Right:} Domain wall structure at a coverage 0.072~\AA$^{-2}$. {\it Bottom Row: Left:}  C7/16 commensurate phase as found in this work (see also Fig.~\ref{Comparison}). {\it Bottom Row: Right:} $^4$He density profile for the incommensurate crystalline phase, at a coverage 0.111~\AA$^{-2}$, which is the maximum first layer coverage. The temperature is $T=0.5$K in all cases.}
\end{figure}

As the chemical potential is increased above $\mu = -130$ K, the average $^4$He density jumps from zero to the value 0.0636 \AA$^{-2}$ corresponding to the crystalline, commensurate   C1/3 phase, wherein one of three equivalent adsorption sites of the lattice is occupied by a single $^4$He atom, as is shown in Fig.~\ref{PicC7/16}. In some simulations a metastable liquid phase with a slightly higher value of the energy per atom was observed, which is typical for a first order transition. Conceivably, such a phase might be seen in experiments.\cite{GreywallHeat} No evidence of a low-coverage thermodynamically stable superfluid (which exists on a smooth, flat substrate\cite{bon99,bon09}) is observed.
\\ \indent
The C1/3 crystal forms at a distance  $z = 2.864(7) \AA$, with a full width at half maximum $\Delta z = 0.65(2) \AA$ (these numbers hold for the entire first layer),  only slightly greater than that for graphite. 
The thermodynamic stability of the C1/3 phase is signaled by an extended plateau of the density as a function of $\mu$, hinting at a large gap of about $6.0(5)$ K in the spectrum and implying a zero superfluid response and zero compressibility. All of this is very similar to hardcore bosons on a lattice with long-range interactions and also to what is observed on graphite.\cite{Corboz} 
However, Ref. \onlinecite{Cazorla} reports a very small (less than 1\%), but finite value  of the superfluid fraction at $T$=0, increasing to approximately 14\% upon doping such phase with vacancies. Assuming a Kosterlitz-Thouless scenario for the U(1) transition for the C1/3 phase, the transition temperature corresponding to the data of Ref. \onlinecite{Cazorla} corresponds to $T \approx 5$mK. In Tables~\ref{table:8x6} and Table~\ref{table:8x8} we computed the superfluid properties of the C1/3 phase down to $T=10$mK for a system consisting of $8 \times 6$ and $8 \times 8$ $^4$He atoms, respectively. We used $200$ times slices per inverse Kelvin, except for the $T=10$mK case on the $8 \times 8$ system where 160 slices per inverse Kelvin were used because of limitations of computer memory. This led to a reduction of less than a percent on the kinetic energy. Simulations were started with an initial superfluid. In Tables~\ref{table:8x6} and~\ref{table:8x8} we see that all data are consistent with insulating behavior and a large gap of a few K in the C1/3 phase, which is consistent with the results of Ref. \onlinecite {Kwon}. Non-zero winding numbers are possible on our small system sizes (ruling out ergodicity problems), but are a result of finite size effects and disappear exponentially with system size. In particular, at $T=10$mK a Kosterlitz-Thouless scenario would have predicted  a strong superfluid response of the order of $50\%$ of the value reported in Ref.~\onlinecite{Cazorla} (at $T=0$K). This is clearly not the case and we conclude that the diffusion Monte Carlo results of Ref.~\onlinecite{Cazorla} are irreproducible, prompting a critical and systematic analysis of the diffusion Monte Carlo methodology (finite system size, influence of the trial wave function and finite population bias\cite{bias}). With respect to the thermodynamic stability of commensurate phases doped with vacancies reported in Refs.~\onlinecite{Cazorla,Kwon}, the use of the grand-canonical ensemble employed here offers distinct advantages over other ensembles to find phase separation.

\begin{figure}
\includegraphics[angle={-90},width=1.0\linewidth]{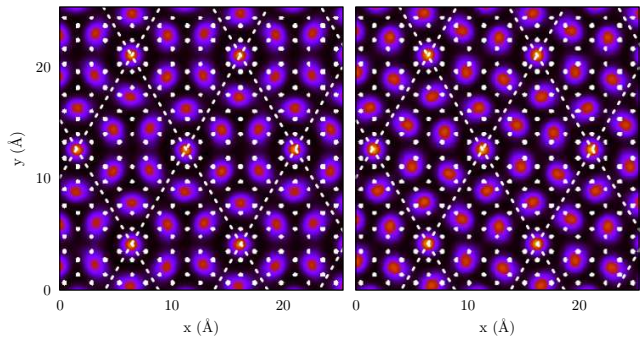}
\caption{\label{Comparison} (Color online). {\it Left:} $^4$He density profile for the C7/16 phase as seen in Ref. \onlinecite{Kwon}. The unit cells are shown with dashed lines. The small (white) dots represent the underlying graphene lattice. The large (red) dots are maxima of the density. 
{\it Right:} $^4$He density profile for the C7/16 phase obtained in this study. }
\end{figure}

Further increasing the chemical potential leads to the appearance of domain walls, as is seen in Fig.~\ref{PicC7/16}, which first occur along one principal axis of the C1/3 phase. This phase is akin to striped phases, but with increasing chemical potential a proliferation of domain walls is observed, along more and more principal axes. In the thermodynamic limit, it has no superfluid response. This process stops when a commensurate crystalline phase labelled as C7/16 with a coverage of 0.0835~\AA$^{-2}$ is formed, and which can be seen as every unit cell of the C1/3 phase being surrounded by domain walls.
The C7/16 structure has a rhombic unit cell, with 7 $^4$He atoms  distributed over 16 absorption sites. For this commensurate structure, too, no finite superfluid signal is computed.
Fig.~\ref{PicC7/16} shows the $^4$He density profile for the C7/16 phase observed in our large scale simulations.
Surprisingly, the helium atoms are in our case distributed more uniformly over the unit cell than in previous works.~\cite{Corboz, Kwon} Moreover, the structure is slightly rotated   (a comparison with the commensurate phases seen by others is shown in Fig.~\ref{Comparison}). We performed  the simulation with a smaller system size (the same as the one used in Ref.~\onlinecite{Kwon}), and we also used  as initial configuration one corresponding to the the density profile obtained in Ref.~\onlinecite{Kwon}, but our Monte Carlo simulation still stabilized, after a sufficiently long time, to the same structure of higher symmetry inside the unit cell,  shown in Fig.~\ref{Comparison}. Thus, we conclude that the difference is not attributable to system size, nor is it a result of a specific choice of initial atomic configuration.
\\ \indent
The most likely reason for the difference between our and previous results is the modeling of the helium-carbon interaction. While we use here a two-body potential ($u$ in Eq. \ref{ham}) with fixed carbon positions, thereby retaining full rotational symmetry, previous studies made use of an external one-body potential obtained by summing anisotropic Lennard-Jones potentials proposed to fit helium scattering from graphite.
The slightly different ways of accounting for corrugation may well be at the root of the lower symmetry structure found in previous studies, and this is supported by comparing the actual potentials felt at locations inside the unit cell, which differ between the two methods.
\\ \indent
On increasing $\mu$ even further, the C7/16 phase is replaced by an incommensurate, compressible solid phase, sketched in Fig.~\ref{PicC7/16} (panel on the left in the bottom row). The density increases linearly with chemical potential until a density of  0.107(3)~\AA$^{-2}$ is reached where the first atoms in the second layer are found. This value is slightly higher than the one reported in Ref.~\onlinecite{Kwon}, but lower than the one from Gordillo {\it et al.}, namely 0.115~\AA$^{-2}$ \cite{Gordillo2nd}. It is also lower than the corresponding one for a graphite substrate, either from theoretical calculations,\cite{Corboz} namely 0.114~\AA$^{-2}$, or inferred from specific heat measurements,\cite{GreywallHeat} i.e., 0.127~\AA$^{-2}$, something that can be attributed to the weaker substrate attraction.

\section{\label{SecondLayer}Results for the second Layer}

\begin{figure}
\includegraphics[angle={-90},width=1.0\linewidth]{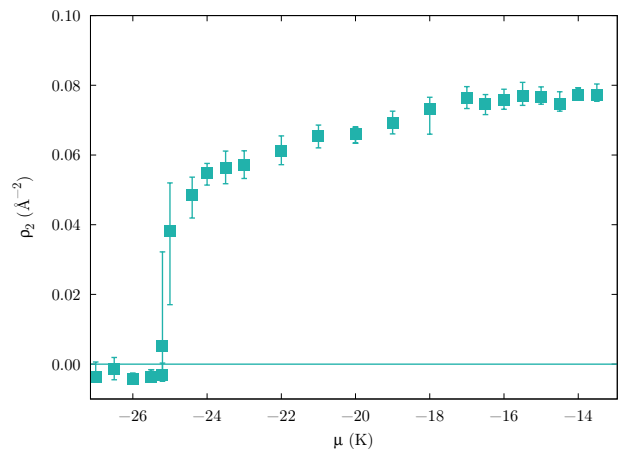}
\caption{\label{2nd}(Color online). Average coverage (offset by the maximum first layer density, $\rho_1 = 0.111$ \AA$^{-2}$) versus chemical potential for the second $^4$He adlayer on graphene at a temperatue $T=1$K with a simulation box of size $29.51\AA \times 34.08\AA$. Data are shown for different simulations, corresponding to different initial atomic arrangements. The second layer becomes populated around $\mu = -25$ K, forming a homogeneous superfluid. The horizontal line denotes the maximum density in the first layer before layer promotion starts; this first layer density was also seen when the second layer is strongly occupied.  For higher $\mu$, a first-order phase transition to an incommensurate solid occures.  No evidence of any supersolid phase is found.}
\end{figure}

\begin{figure}
\includegraphics[angle={-90},width=1.0\linewidth]{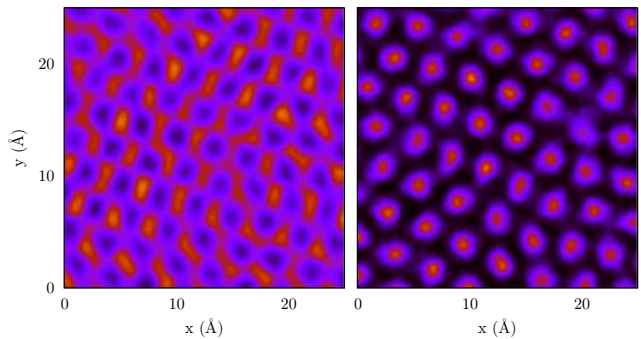}
\caption{\label{Pic2nd} (Color online). {\it Left:}  second layer density profile at a two-dimensional density  0.0731~\AA$^{-2}$ and temperature $T=1$K. The dark spots correspond to the underlying helium first layer atoms. {\it Right:} second layer density profile for the incommensurate phase, at a two-dimensional  density 0.0781~\AA$^{-2}$, which is immediately below third layer promotion.}
\end{figure}

We now turn to the second layer. The first stable phase that we observe is a uniform liquid, which is superfluid at the temperature considered here (Fig.~\ref{2nd} and Fig.~\ref{Pic2nd}).  The equilibrium coverage is roughly 0.150 \AA$^{-2}$, consistent with the two-dimensional density of the superfluid being close to the equilibrium density of two-dimensional $^4$He, {\it i.e.} 0.043 \AA$^{-2}$.
This superfluid phase has a relatively wide domain of existence, extending up to a layer density of $0.076~\pm 0.003$ \AA$^{-2}$. Within the quoted  statistical uncertainty, this value is in agreement with that of Ref. \onlinecite{Gordillo2nd}, and also consistent with that  found on graphite,\cite{Corboz} namely  0.076~\AA$^{-2}$.\\ 
\indent
On raising the chemical potential, a first-order phase transition to an incommensurate crystal is observed, which persists up to atomic third layer promotion. The density profile of the incommensurate solid phase is shown in Fig.~\ref{Pic2nd}. Its superfluid response is zero in the thermodynamic limit. This is qualitatively the same behavior observed for graphite (see Ref. \onlinecite{Corboz}). Nowhere in the second layer are commensurate phases found, and in particular we find no evidence of any ``supersolid" phase. The arguments for the absence of the suggested C7/12 and C4/7 phases formulated in Ref.~\onlinecite{Corboz} for graphite apply equally well for graphene. 
Gordillo and Boronat suggested that there could be a coexistence region between liquid and incommensurate phase.\cite{Gordillo2nd} Since the (quantum) transition between a liquid and solid where the substrate does not play a role has to be first order, such a region must be sought around $\mu = -18$K in Fig.~\ref{2nd}. We did not analyze this first order transition in more detail, since it was very narrow already in graphite.~\cite{Corboz}
As a final remark, we note that the $^4$He density between the layers is essentially zero, {\it i.e.},  layers constitute distinct, effectively two-dimensional systems.


\section{\label{Conclusion}Conclusion}

We studied the phase diagram of the first and second layer of $^4$He adsorbed on graphene. In the first layer, we find a very stable C1/3 phase, followed by a domain wall phase, a C7/16 phase and an incommensurate solid. This is the same as for graphite,\cite{Corboz} and in line with previous studies.\cite{Kwon, Gordillo1st} The only difference we observed consisted of the density distribution of the helium atoms inside the C7/16 unit cell, where we observed a small rotation (and more symmetric) distribution. We attribute this difference to the modeling of the carbon-helium interaction, which was done slightly differently than in previous studies: We used a fully rotationally symmetric two-body potential between carbon and helium atoms, and kept a (classical) hexagonal carbon lattice in the simulation, while previous studies used a one-body potential that could be isotropic or anisotropic. We neglected defects, disorder,  and other lattice imperfections in the graphene.  For the second layer, our results are also in line with the ones found on graphite~\cite{Corboz} and previous studies showing a superfluid and an incommensurate solid at higher coverages. No commensurate structures are found, nor any supersolids. Layer promotion occurs at a slightly lower chemical potential than for graphite, because of the slightly weaker helium-substrate potential.

In future work, we will study more realistic descriptions of graphene, which is not entirely flat but has intrinsic microscopic roughening. Suspended graphene sheets under a transmission electron microscope were seen to have a surface normal varying by several degrees and the out of plane deformations reach 1 nm. \cite{MeyerGraphene} 
This roughened structure is thought to suppress and stabilize thermal vibrations and to influence the behavior as an adsorbent. Our setup can also be modified to study helium inside carbon nano-tubes: instead of using periodic boundary conditions in both $x$ and $y$ directions on the graphene substrate, one could use periodic boundary conditions only in the $x$ direction.

{\it Acknowledgements --} We wish to thank A. B. Kuklov for fruitful discussions. MB thanks the Arnold Sommerfeld Center at the LMU Munich for hospitality during the early stages of this work. LP acknowledges support from Marie Curie CIG Grant "FDIAGMC" (PCIG11-GA-2012-321918).

\end{document}